\documentclass[preprint,3p]{elsarticle}
\pdfoutput=1

\usepackage{amsfonts}
\usepackage{amsmath}
\usepackage{amssymb}
\usepackage{graphicx}
\usepackage{subfigure}
\usepackage{units}
\usepackage{multirow}
\usepackage{color}
\usepackage{amsmath}
\usepackage{textcomp}
\usepackage{bm}
\usepackage[colorlinks=true,linkcolor=blue]{hyperref}
\usepackage{multicol}
\usepackage{slashbox}

\biboptions{numbers,sort&compress}

\RequirePackage{lineno}

\newcommand{\ee}{e^+e^-}
\newcommand{\dzero}{D^0}
\newcommand{\dbarzero}{\overline{D}{}^{0}}

\newcommand{\gev}{\,\unit{GeV}}
\newcommand{\gevc}{\,\unit{GeV}/\emph{c}}
\newcommand{\gevcc}{\,\unit{GeV}/\emph{c}^2}
\newcommand{\br}[1]{\mathcal{B}_{#1}}
\newcommand{\srd}{r}
\newcommand{\kpi}{{K^-\pi^+}}

\newcommand{\strph}{\delta_{K\pi}}
\newcommand{\mbc}{M_\mathrm{BC}}
\newcommand{\delE}{\Delta\emph{E}}

\newcommand {\ResultOfAcp}{(12.7\pm1.3\pm0.7)\times10^{-2}}
\newcommand {\ResultOfCosStrPh}{1.02\pm0.11\pm0.06\pm0.01}

\begin{document}
\setpagewiselinenumbers
\modulolinenumbers[2]

\begin{frontmatter}

\title{Measurement of the $D\to K^-\pi^+$ strong phase difference in $\psi(3770)\to D^0\overline{D}{}^0$}

\author{
\small
M.~Ablikim$^{1}$, M.~N.~Achasov$^{8,a}$, X.~C.~Ai$^{1}$, O.~Albayrak$^{4}$, M.~Albrecht$^{3}$, D.~J.~Ambrose$^{41}$, F.~F.~An$^{1}$, Q.~An$^{42}$, J.~Z.~Bai$^{1}$, R.~Baldini Ferroli$^{19A}$
% ...
\\
\vspace{0.2cm}
(BESIII Collaboration)\\
\vspace{0.2cm}
{\it
$^{1}$ Institute of High Energy Physics, Beijing 100049, People's Republic of China\\
% ...
$^{3}$ Bochum Ruhr-University, D-44780 Bochum, Germany\\
$^{4}$ Carnegie Mellon University, Pittsburgh, Pennsylvania 15213, USA\\
% ...
$^{8}$ G.I. Budker Institute of Nuclear Physics SB RAS (BINP), Novosibirsk 630090, Russia\\
% ...
$^{19}$ (A)INFN Laboratori Nazionali di Frascati, I-00044, Frascati, Italy; (B)INFN and University of Perugia, I-06100, Perugia, Italy\\
% ...
$^{41}$ University of Rochester, Rochester, New York 14627, USA\\
$^{42}$ University of Science and Technology of China, Hefei 230026, People's Republic of China\\
% ...
$^{a}$ Also at the Novosibirsk State University, Novosibirsk, 630090, Russia\\
% ...
}
\vspace{0.4cm}
}


\begin{abstract}

We study $D^0\overline{D}{}^0$ pairs produced in $e^+e^-$ collisions at
$\sqrt{s}=3.773$\,GeV using a data sample of 2.92 fb$^{-1}$ collected with the BESIII
detector.
 We measure the asymmetry $\mathcal{A}^{CP}_{K\pi}$ of the
branching fractions of $D \to K^-\pi^+$ in $CP$-odd and $CP$-even
eigenstates to be $(12.7\pm1.3\pm0.7)\times10^{-2}$. $\mathcal{A}^{CP}_{K\pi}$ can be
used to extract the strong phase difference $\delta_{K\pi}$ between
the doubly Cabibbo-suppressed process $\overline{D}{}^{0}\to K^-
\pi^+$ and the Cabibbo-favored process $D^0\to K^- \pi^+$. Using
world-average values of external parameters, we obtain
$\cos\delta_{K\pi} = 1.02\pm0.11\pm0.06\pm0.01$.
Here, the first and second uncertainties are statistical and
systematic, respectively, while the third uncertainty arises from the
external parameters. This is the most precise measurement of
$\delta_{K\pi}$ to date.

\end{abstract}


\end{frontmatter}


\section{Introduction}\label{sec:intro}

Within the Standard Model, the short-distance
  contribution to $\dzero$-$\dbarzero$ oscillations is highly
  suppressed by the GIM mechanism~\cite{GIM} and by the magnitude of
  the CKM matrix elements~\cite{CKM} involved.
However, long distance effects, which cannot be reliably calculated,
will also affect the size of mixing. Studies of $\dzero$-$\dbarzero$
oscillation provide knowledge of the size of these long-distance
effects and, given improved calculations, can contribute to searches
for new physics~\cite{Bigi}. In addition, improved constraints on
charm mixing are important for studies of $CP$ violation ($CPV$) in
charm physics.

Charm mixing is described by two dimensionless parameters
$$x=2\frac{M_1-M_2}{\Gamma_1+\Gamma_2}\label{1} \hspace{15pt}    y=\frac{\Gamma_1-\Gamma_2}{\Gamma_1+\Gamma_2},$$
where $M_{1,2}$ and $\Gamma_{1,2}$ are the masses and widths of the two
mass eigenstates in the $D^0$-$\overline{D^0}$ system.
The most precise determination of the
mixing parameters comes from the measurement of the time-dependent
decay rate of the wrong-sign process $D^0\to K^+\pi^-$. These analyses
are sensitive to $y'\equiv y\cos\strph-x\sin\strph$ and $x'\equiv
x\cos\strph +y\sin\strph$~\cite{yp1}, where $\strph$ is the strong
phase difference between the doubly Cabibbo-suppressed (DCS) amplitude
for $\dbarzero\to K^-\pi^+$ and the corresponding Cabibbo-favored (CF)
amplitude for $\dzero\to K^-\pi^+$.  In particular,
 \begin{eqnarray}
     \frac{\langle K^-\pi^+|\dbarzero\rangle}{\langle K^-\pi^+|\dzero\rangle}
    = -\srd e^{-i\strph},\label{5}
 \end{eqnarray}
where
\begin{equation}
    \srd=\left|\frac{\langle K^-\pi^+|\dbarzero\rangle}{\langle K^-\pi^+|\dzero\rangle}\right|.\label{6}  \nonumber
\end{equation}
Knowledge of $\strph$ is important for extracting $x$ and $y$ from
$x'$ and $y'$. In addition, a more accurate $\strph$ contributes to
precision determinations of the CKM unitarity angle
$\phi_3$\footnote{$\gamma$ is also used in the literature.} via the ADS method~\cite{ADS}.

Using quantum-correlated techniques, $\strph$ can be accessed in the
mass-threshold production process $\ee\to
\dzero\dbarzero$~\cite{Cheng:2007uj}. In this process, $\dzero$ and
$\dbarzero$ are in a $C$-odd  quantum-coherent state where the two
mesons necessarily have opposite $CP$ eigenvalues~\cite{Bigi}. Thus,
threshold production provides a unique way to identify the $CP$ of one
neutral $D$ by probing the decay of the partner $D$. Because $CPV$ in
$D$ decays is very small compared with the mixing parameters, we will
assume no $CPV$ in our analysis. In this paper, we often refer to
$K^-\pi^+$ only for simplicity, but charge-conjugate modes are always
implied when appropriate.

We denote the asymmetry of $CP$-tagged $D$ decay rates to $\kpi$ as
\begin{eqnarray}
\mathcal{A}^{CP}_{K\pi}\equiv\frac{\br{D^{S-}\to \kpi}-\br{D^{S+}\to \kpi}}{\br{D^{S-}\to \kpi}+\br{D^{S+}\to \kpi}},\label{12}
\end{eqnarray}
where $S+$ ($S-$) denotes the $CP$-even ($CP$-odd) eigenstate.
To lowest order in the mixing parameters, we have the relation~\cite{Xing:1996pn,CLEO-c2}
\begin{eqnarray}
   2\srd \cos\strph + y &=& (1+R_{\mathrm{WS}})\cdot\mathcal{A}^{CP}_{K\pi},
    \label{11}
\end{eqnarray}
where $R_\mathrm{WS}$ is the decay rate ratio of the wrong sign
process $\dbarzero\to\kpi$ (including the DCS~decay and $D$ mixing
followed by the CF~decay) and the right sign process $\dzero\to\kpi$
($i.e.$, the CF~decay). Here, $\dzero$ or $\dbarzero$ refers to the state at production.
Using external values for the parameters $r$, $y$, and $R_{\mathrm{WS}}$, we can extract $\strph$ from $\mathcal{A}_{CP\to K\pi}$.

We use the $D$-tagging method~\cite{Baltrusaitis:1985iw} to obtain the
branching fractions $\br{D^{S\pm}\to K^-\pi^+}$ as
\begin{equation}
    \br{D^{S\pm}\to K^-\pi^+} = \frac{n_{K^-\pi^+,S\pm}}{n_{S\pm}}\cdot\frac{\varepsilon_{S\pm}}{\varepsilon_{K^-\pi^+,S\pm}}.\label{22}
\end{equation}
Here, $n_{S\pm}$ and $\varepsilon_{S\pm}$ are yields and detection
efficiencies of single tags (ST) of $S\pm$ final states, while
$n_{\kpi,S\pm}$ and $\varepsilon_{\kpi,S\pm}$ are yields and
efficiencies of double tags (DT) of ($S\pm$, $\kpi$) final states,
respectively. Based on an 818\,pb$^{-1}$ data sample collected with
the CLEO-c detector at $\sqrt{s}=3.77\gev$ and a more complex analysis
technique, the CLEO collaboration obtained
$\cos\strph=0.81^{+0.22+0.07}_{-0.18-0.05}$~\cite{CLEO-c2}. Using a
global fit method including external inputs for mixing parameters,
CLEO obtained $\cos\strph=1.15^{+0.19+0.00}_{-0.17-0.08}$~\cite{CLEO-c2}.

In this paper, we present a measurement of $\strph$, using the
quantum correlated productions of $\dzero$-$\dbarzero$ mesons at
$\sqrt{s}=3.773\gev$ in $e^+e^-$ collisions with an integrated
luminosity of 2.92\,fb$^{-1}$~\cite{jiangll} collected with the BESIII
detector~\cite{:2009vd}.

\section{The BESIII Detector}\label{sec:BESIII}

The Beijing Spectrometer (BESIII) views  $\ee$ collisions in the
double-ring collider BEPCII. BESIII is a general-purpose
detector~\cite{:2009vd} with 93\,\% coverage of the full solid angle. From
the interaction point (IP) to the outside, BESIII is equipped with a
main drift chamber (MDC) consisting of 43 layers of drift cells, a
time-of-flight (TOF) counter with double-layer scintillator in the
barrel part and single-layer scintillator in the end-cap part, an
electromagnetic calorimeter (EMC) composed of 6240 CsI(Tl) crystals, a
superconducting solenoid magnet providing a magnetic field of 1.0\,T
along the beam direction, and a muon counter containing multi-layer
resistive plate chambers installed in the steel
  flux-return yoke of the magnet. The MDC spatial resolution is about
135\,$\mu$m and the momentum resolution is about 0.5\,\% for a charged
track with transverse momentum of 1\gevc. The energy resolution for
showers in the EMC is 2.5\,\% at 1\gev.  More details of the
spectrometer can be found in Ref.~\cite{:2009vd}.

\section{MC Simulation}\label{sec:MC}

Monte Carlo (MC) simulation serves to estimate the detection efficiency
and to understand background components. MC samples corresponding to about 10 times
the luminosity of data are generated with a {\sc geant4}-based~\cite{geant4} software
package~\cite{Deng:2007zzb}, which includes simulations of the
geometry of the spectrometer and interactions of particles with the
detector materials. {\sc kkmc} is used to model the beam energy spread
and the initial-state radiation (ISR) in the $\ee$
annihilations~\cite{Jadach:2000ir}. The inclusive MC samples consist
of the production of $D\overline{D}$ pairs with consideration of
quantum coherence for all modes relevant to this analysis, the
non-$D\overline{D}$ decays of $\psi(3770)$, the ISR production of low
mass $\psi$ states, and QED and $q\bar{q}$ continuum processes.
Known decays recorded in the Particle Data Group (PDG)~\cite{PDG} are
simulated with {\sc evtgen}~\cite{evtgen} and the unknown decays with
{\sc lundcharm}~\cite{lund}. The final-state radiation (FSR) off
charged tracks is taken into account with the {\sc photos}
package~\cite{photos}.
MC samples of $D\to S\pm, \overline{D}\to X$ ($X$ denotes inclusive
decay products) processes are used to estimate the ST efficiencies, and
MC samples of $D\to S\pm, \overline{D}\to K\pi$ processes are used to
estimate the DT efficiencies.

\section{Data Analysis}\label{sec:ana}

\begin{table*}[tbp!]
\caption{$D$ decay modes used in this analysis.}{\label{CP_Mode}}
\begin{center}
\begin{tabular}{lc }
\hline
\hline
Type        & Mode  \\
\hline
Flavored    & $K^-\pi^+, K^+\pi^-$  \\
$S+$         & $K^+K^-, \pi^+\pi^-, K^0_S\pi^0\pi^0, \pi^0\pi^0, \rho^0\pi^0$  \\
$S-$         & $K^0_S\pi^0, K^0_S\eta, K^0_S\omega$  \\
\hline
\hline
\end{tabular}

\end{center}
\end{table*}

The decay modes used for tagging the $CP$ eigenstates are listed in
Table~\ref{CP_Mode}, where $\pi^0\to\gamma\gamma$,
$\eta\to\gamma\gamma$, $K^0_S\to\pi^+\pi^-$ and
$\omega\to\pi^+\pi^-\pi^0$.  For each mode, $D$ candidates are
reconstructed from all possible combinations of final-state particles,
according to the following selection criteria.

Momenta and impact parameters of charged tracks are measured by the
MDC. Charged tracks are required to satisfy $|\cos\theta|<0.93$, where
$\theta$ is the polar angle with respect to the beam axis, and have a
closest approach to the IP within $\pm10$\,cm along the beam direction
and within $\pm1$\,cm in the plane perpendicular to the beam
axis. Particle identification is implemented by combining the
information of normalized energy deposition (d$E$/d$x$) in the MDC and
the flight time measurements from the TOF. For a charged $\pi$($K$)
candidate, the probability of the $\pi$($K$) hypothesis is required to
be larger than that of the $K$($\pi$) hypothesis.

Photons are reconstructed as energy deposition clusters in the
EMC. The energies of photon candidates must be larger than 25\,MeV for
$|\cos\theta|<0.8$ (barrel) and 50\,MeV for $0.84<|\cos\theta|<0.92$
(end-cap). To suppress fake photons due to electronic noise or beam
backgrounds, the shower time must be less than 700\,ns
from the event start time~\cite{t0}.
However, in the case that no charged track is detected, the event
start time is not reliable, and instead the shower time
must be within $\pm$500\,ns from the time of the most energetic shower.

Our $\pi^0$ and $\eta$ candidates are selected from pairs of photons
with the requirement that at least one photon candidate
reconstructed in the barrel is used. The mass
windows imposed are $0.115\gevcc<m_{\gamma\gamma}<0.150\gevcc$ for
$\pi^0$ candidates and $0.505\gevcc<m_{\gamma\gamma}<0.570\gevcc$ for
$\eta$ candidates. We further constrain the invariant mass of each
photon pair to the nominal $\pi^0$ or $\eta$ mass, and update the four
momentum of the candidate according to the fit results.

The $K^0_S$ candidates are reconstructed via $K^0_S\to\pi^{+}\pi^{-}$
using a vertex-constrained fit to all pairs of oppositely charged
tracks, with no particle identification requirements. These tracks
have a looser IP requirement: their closest approach to the IP is
required to be less
than 20\,cm along the beam direction, with no requirement in the
transverse plane. The $\chi^2$ of the vertex fit is required to be
less than 100. In addition, a second fit is performed,
constraining the $K^0_S$ momentum to point back to the IP. The flight
length, $L$, obtained from this fit must satisfy $L/\sigma_L>2$, where
$\sigma_L$ is the estimated error on $L$. Finally, the invariant mass
of the $\pi^{+}\pi^{-}$ pair is required to be within $(0.487,
0.511)\gevcc$, which corresponds to
three times the experimental mass resolution.

\subsection{Single tags using $CP$ modes}\label{sec:single}

For the $CP$-even and $CP$-odd modes, the two variables beam-constrained
mass $\mbc$ and energy difference $\delE$ are used to identify the
signals, defined as follows:
\begin{equation}
   \mbc\equiv\sqrt{E^2_{\unit{beam}}/c^4-|\vec{p}_{D}|^2/c^2},\label{25} \nonumber
\end{equation}
\begin{equation}
    \delE\equiv E_{D}-E_{\unit{beam}}.\label{26} \nonumber
\end{equation}
Here $\vec{p}_{D}$ and $E_{D}$ are the total momentum and energy of
the $D$ candidate, and $E_{\unit{beam}}$ is the beam energy. Signals
peak around the nominal $D$ mass in $\mbc$ and around zero in
$\delE$. Boundaries of $\delE$ requirements are set at approximately
$\pm3\sigma$, except that those of modes containing a $\pi^0$ are set as
($-4\sigma, +3.5\sigma$) due to the asymmetric distributions. In each
event, only the combination of $D$ candidates with the least $|\delE|$
is kept per mode.

\begin{table*}[tbp!]
\caption{Requirements on $\Delta E$ for different $D$ reconstruction modes.}{\label{deltaE}}
\begin{center}
\begin{tabular}{lc }
\hline
\hline
Mode                    & Requirement (GeV) \\
\hline
$K^+K^-$                & $-0.025<\Delta E<0.025$ \\
$\pi^+\pi^-$            & $-0.030<\Delta E<0.030$ \\
$K^0_S\pi^0\pi^0$       & $-0.080<\Delta E<0.045$ \\
$\pi^0\pi^0$            & $-0.080<\Delta E<0.040$ \\
$\rho^0\pi^0$           & $-0.070<\Delta E<0.040$ \\
$K^0_S\pi^0$            & $-0.070<\Delta E<0.040$ \\
$K^0_S\eta$             & $-0.040<\Delta E<0.040$ \\
$K^0_S\omega$           & $-0.050<\Delta E<0.030$ \\
$K^{\pm}\pi^{\mp}$      & $-0.030<\Delta E<0.030$ \\
\hline
\hline
\end{tabular}

\end{center}
\end{table*}

\begin{figure*}[thbp!]
\centering
\includegraphics[width=0.95\linewidth]{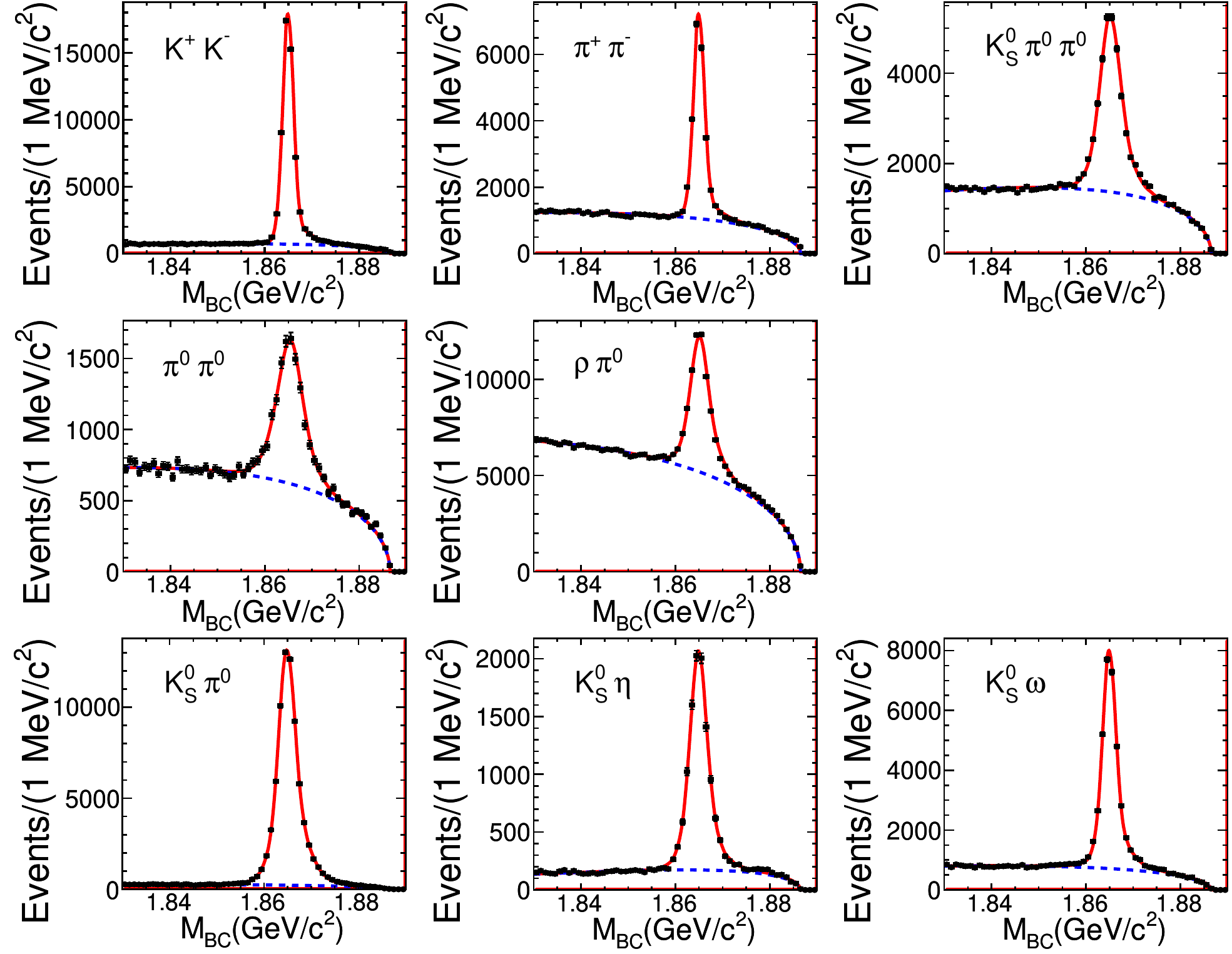}
\caption{The $\mbc$ distributions of the single-tag (ST) $CP$
  modes. Data are shown as points with error bars. The solid lines are
  the total fits and the dashed lines are the background
  contribution.}{\label{ST_all}}
\centering
\end{figure*}

\begin{table*}[tp!]
\caption{Yields and efficiencies of all single-tag (ST) and double-tag
  (DT) modes.  First, we list the ST ($CP$ mode) yields ($n_{S\pm}$)
  and corresponding efficiencies ($\varepsilon_{S\pm}$) and then the
  DT mode yields ($n_{K\pi,S\pm}$) and efficiencies
  ($\varepsilon_{K\pi,S\pm}$).  Uncertainties are statistical
  only.}{\label{ST_DT}}
\centering
\begin{tabular}{lcc}
\hline
\hline
ST Mode         & $n_{S\pm}$          & $\varepsilon_{S\pm}$(\%)   \\
$K^+K^-$                & $56156\pm261$     & $62.99\pm0.26$ \\
$\pi^+\pi^-$            & $20222\pm187$     & $65.58\pm0.26$\\
$K^0_S\pi^0\pi^0$       & $25156\pm235$     & $16.46\pm0.07$ \\
$\pi^0\pi^0$            & $7610\pm156$      & $42.77\pm0.21$ \\
$\rho\pi^0$             & $41117\pm354$    & $36.22\pm0.21$ \\
$K^0_S\pi^0$            & $72710\pm291$     &  $41.95\pm0.21$ \\
$K^0_S\eta$             & $10046\pm121$     & $35.12\pm0.20$  \\
$K^0_S\omega$           & $31422\pm215$     & $17.88\pm0.10$  \\
\hline
DT Mode      & $n_{K\pi,S\pm}$        & $\varepsilon_{K\pi,S\pm}$(\%)   \\
$K\pi,~K^+K^-$               & $1671\pm41$         & $42.33\pm0.21$ \\
$K\pi,~\pi^+\pi^-$           & $610\pm25$          & $44.02\pm0.21$ \\
$K\pi,~K^0_S\pi^0\pi^0$      & $806\pm29$          & $12.86\pm0.13$\\
$K\pi,~\pi^0\pi^0$           & $213\pm14$          & $30.42\pm0.18$ \\
$K\pi,~\rho\pi^0$            & $1240\pm35$         & $25.48\pm0.16$ \\
$K\pi,~K^0_S\pi^0$           & $1689\pm41$         & $29.06\pm0.17$ \\
$K\pi,~K^0_S\eta$            & $230\pm15$          & $24.84\pm0.16$ \\
$K\pi,~K^0_S\omega$          & $747\pm27$          & $12.60\pm0.06$ \\
\hline\hline
\end{tabular}

\end{table*}

In the $K^+K^-$ and $\pi^+\pi^-$ modes, backgrounds of cosmic rays and
Bhabha events are removed with the following requirements.  First, the
two charged tracks used as the $CP$ tag must have a TOF time
difference less than 5\,ns and they must not be consistent with being
a muon pair or an electron-positron pair.  Second, there must be at
least one EMC shower (other than those from the $CP$ tag tracks) with
an energy larger than 50\,MeV or at least one additional charged track
detected in the MDC. In the $K^0_S\pi^0$ mode, backgrounds due to
$D^0\to\rho\pi$ are negligible after restricting the decay length of
$K^0_S$ with $L/\sigma_L>2$. In the $\rho^0\pi^0$ and  $K^0_S\omega$
modes, mass ranges of $0.60\gevcc<m_{\pi^+\pi^-}<0.95\gevcc$ and
$0.72\gevcc<m_{\pi^+\pi^-\pi^0}<0.84\gevcc$ are required for
identifying $\rho$ and $\omega$ candidates, respectively.

After applying the criteria on $\delE$ in Table~\ref{deltaE} in all
the $CP$ modes, we plot their $\mbc$ distributions in
Fig.~\ref{ST_all}, where the peaks at the nominal $D^0$ mass are
evident. Maximum likelihood fits to the events in Fig.~\ref{ST_all}
are performed, where in each mode the signals are modeled with the
reconstructed signal shape in MC simulation convoluted with a smearing
Gaussian function, and backgrounds are modeled with the ARGUS
function~\cite{Argus}. The Gaussian functions are supposed to
compensate for the resolution differences between data and MC
simulation. Based on the fit results, the estimated yields of the $CP$
modes are given in Table~\ref{ST_DT}, along with their MC-determined
detection efficiencies.

\subsection{Double tags of the $\kpi$ and $CP$ modes}\label{double}

\begin{figure*}[tp!]
\begin{center}
\includegraphics[width=0.9\linewidth]{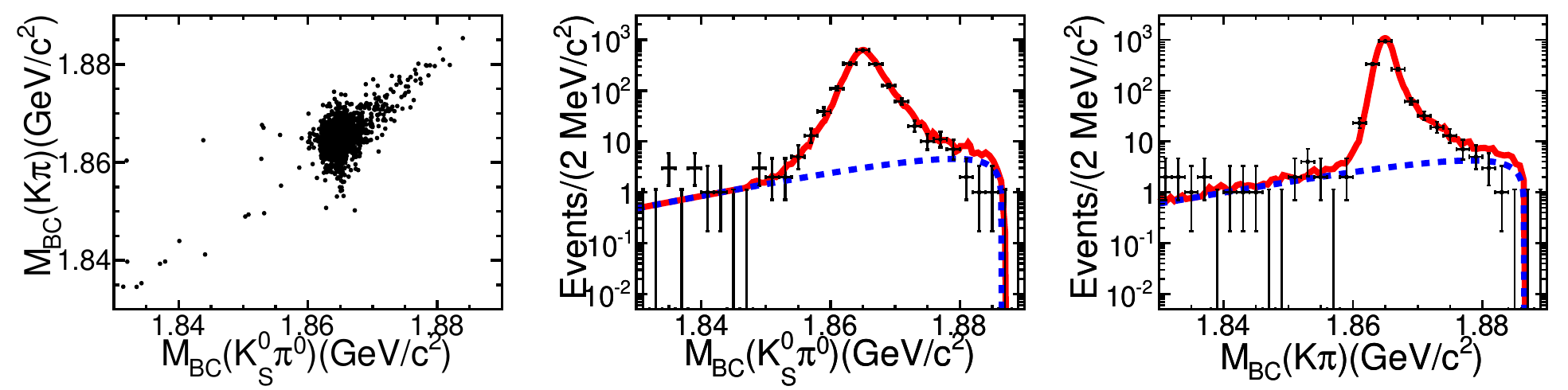}
\caption{An illustration of our DT yield analysis, using the $K\pi$,
  $K_S^0\pi^0$ mode.  A scatter plot (left) of the two $\mbc$ values
  is displayed, along with projections of the two-dimensional fit to
  the same data (middle and right). The solid lines are the total fits
  and the dashed lines are the background
  contribution.}{\label{DT_Kpi}}
\end{center}
\end{figure*}

In the surviving ST $CP$ modes, we reconstruct $D\to\kpi$ among the
unused charged tracks. The $D\to\kpi$ candidate must pass the $\delE$
requirement listed in Table~\ref{deltaE}; in the case of multiple
candidates, the one with the smallest $|\delE|$ is chosen. The DT
signals peak at the nominal $D^0$ mass in both $\mbc(S\pm)$  and
$\mbc(K\pi)$.
To extract the signal yields, two-dimensional maximum likelihood fits
to the distributions of $\mbc(S\pm)$ vs. $\mbc(K\pi)$ are
performed. The signal shapes are derived from MC simulations, and the
background shapes contain continuum background and mis-partitioning
background where some final-state particles are interchanged between
the $\dzero$ and $\dbarzero$ candidates in the reconstruction
process. Figure~\ref{DT_Kpi} shows an example of the results for one
sample DT combination, ($K\pi$, $K_S^0\pi^0$). Table~\ref{ST_DT} lists
the yields of the DT modes and their corresponding detection
efficiencies as determined with MC simulations.

\section{Purities of the $CP$ Modes}\label{CPdouble}

It is necessary to determine the $CP$-purity of our ST modes. For the
$K_S^0\pi^0$ $(K_S^0\eta)$ mode, the issue is the background under
the $K^0_S$ peak. We use the sideband regions of the $K^0_S$ mass,
[0.470, 0.477]$\gevcc$ and [0.521, 0.528]$\gevcc$, in the
$m_{\pi^+\pi^-}$ distributions, to estimate the backgrounds from
$\pi^+\pi^-\pi^0$ $(\pi^+\pi^-\eta)$. 
The purity is estimated to be 98.5\% (almost 100\%) for the $K_S^0\pi^0$ $(K_S^0\eta)$ mode.
For the $K_S^0\omega$,
$K^0_S\pi^0\pi^0$ and $\rho^0\pi^0$ modes, due to the complexity of
the involved non-resonant and resonant processes, we evaluate the
$CP$-purity directly from our data.  We use additional DT
combinations, with a clean $CP$-tag in combination with the mode we
wish to study.  We look for signals where both $D$ mesons decay with
equal $CP$ eigenvalue. If $CP$ is conserved, the same-$CP$ process is
prohibited in the quantum-correlated $D\overline{D}$ production at
threshold, unless our studied $CP$ modes are not pure.
If we take $f_S$ as the fraction of the right $CP$ components in the $CP$ tag mode, we have
the yields of the same-$CP$ process written as
\begin{equation}
    n_{S',S}=(1-f_S)\cdot n_S \cdot \br{D\to S'}\cdot \varepsilon_{S',S}/\varepsilon_{S},\nonumber
\end{equation}
where mode $S'$ is chosen to be (nearly) pure in its $CP$ eigenstate.

\begin{figure*}[tp!]
\begin{center}
\includegraphics[width=0.9\linewidth]{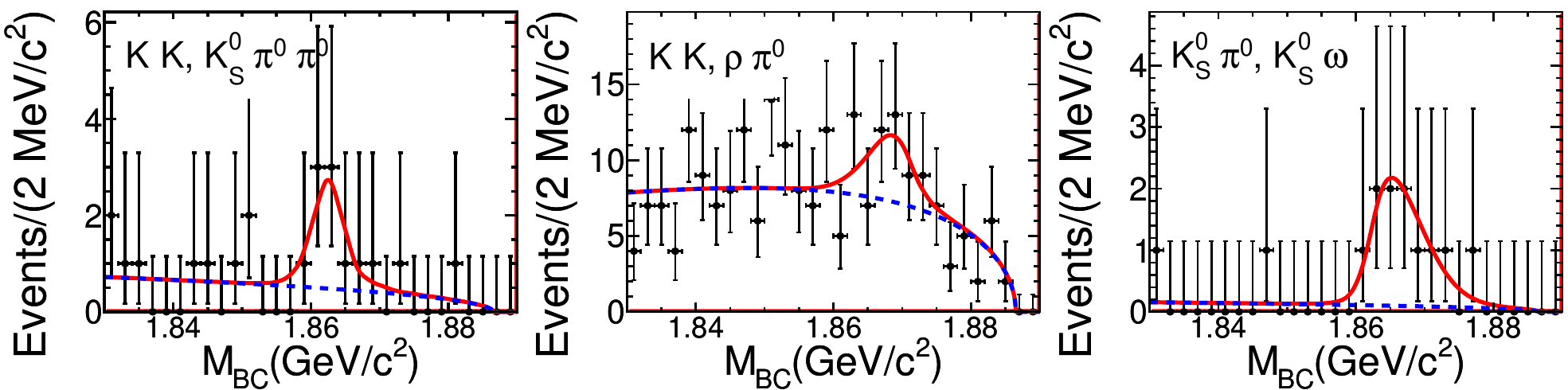}
\caption{The $\mbc$ distributions from our $CP$-purity tests using
  same-$CP$ processes $(S',S)$, with fits to the total (solid) and
  background (dashed) contributions. Both $S$ and $S'$ are $CP$
  eigenstates of $D$ decays.}{\label{DT_all}}
\end{center}
\end{figure*}

We take the modes $K_S^0\pi^0$ ($S'-$) and $K^+K^-$ ($S'+$) as our
clean $CP$ tags to test the $S-$ and $S+$ purities of our ST modes,
respectively.
We analyze our data to find $(S', S)$ events using selection
criteria similar to those described in Sec.~\ref{double}. However, a
simplified procedure is used to obtain the yields. We implement a
one-dimensional fit  to the $\mbc(S)$ distributions for the signal
mode $S$ of interest, while restricting the  $\mbc(S')$ distributions
for the tagging modes $S'$ in the signal region
$1.860\gevcc<\mbc(K^+K^-)<1.875\gevcc$ and
$1.855\gevcc<\mbc(K^0_S\pi^0)<1.880\gevcc$.
The DT signals are described with the signal MC shape convoluted with
a Gaussian function, and backgrounds are modeled with the ARGUS
function. Figure~\ref{DT_all} shows the $\mbc(S)$ distributions in the
DT events and the fits to the distributions. Table~\ref{CP_purity}
lists the DT yields and the corresponding detection efficiencies. In
the tested $CP$ modes, the observed numbers of the same-$CP$ events are
quite small and nearly consistent with zero, which indicates that
$f_S$ is close to 1. This one-dimensional fit may let certain peaking
backgrounds survive; however, an over-estimated $n_{S',S}$ leads to a
more conservative evaluation of $f_S$.

\begin{table*}[tbp!]
\caption{The same-$CP$ yields and the corresponding efficiencies used in our
$CP$-purity tests. The uncertainties are statistical only. The last
column presents the obtained $f_S$ and numbers in the parentheses are
the lower limits of the $f_S$ at 90\% confidence
level.}{\label{CP_purity}}
\begin{center}
\begin{tabular}{lccc}
\hline
\hline
Mode ($S',S$)                  & $n_{S',S}$     & $\varepsilon_{S',S}$(\%)  & $f_S$(\%) \\
\hline
$K^+K^-,~K^0_S\pi^0\pi^0$      & $8\pm3$        & $11.80\pm0.11$            & $91.6\pm16.7$  $(>86.8)$ \\
$K^+K^-,~\rho^0\pi^0$          & $13\pm8$       & $24.44\pm0.16$            & $84.0\pm12.6$  $(>70.6)$ \\
$K^0_S\pi^0,~K^0_S\omega$      & $7\pm3$        & $6.77\pm0.08$             & $94.6\pm8.0$  $(>90.6)$ \\
\hline
\hline
\end{tabular}

\end{center}
\end{table*}

\section{Systematic Uncertainties}\label{sysu}

\begin{table*}[tp!]
\caption{A summary of mode-dependent fractional systematic uncertainties, in percent. A ``--'' means the systematic uncertainty is negligible. }{\label{uncor_sys}}
\begin{center}
\begin{tabular}{lcccccccc}
\hline
\hline
Source              &$K^+K^-$   &$\pi^+\pi^-$   &$K^0_S\pi^0\pi^0$   &$\pi^0\pi^0$   &$\rho^0\pi^0$   &$K^0_S\pi^0$   &$K^0_S\eta$   & $K^0_S\omega$ \\
\hline
$\delE$ requirement & 0.6     & 0.5      & 0.9       & 0.7     & 1.8    & 0.7   & 0.5     & 1.5   \\
Fitting             & 0.9     & 1.0      & 1.5       & 1.7     & 0.2    & 0.1   & 0.8     & 2.0   \\
$CP$ purity         & --      & --       & 1.8       & --      & 3.5    & 0.6   & --      & 1.2  \\
\hline
Quadratic sum       & 1.1     & 1.1      & 2.5       & 1.8     & 3.9    & 0.9   & 0.9     & 2.8  \\
\hline
\hline
\end{tabular}

\end{center}
\end{table*}

In calculating $\mathcal{A}^{CP}_{K\pi}$, uncertainties of most of
efficiencies cancel out, such as those for tracking, particle
identification and $\pi^0$/$\eta$/$K^0_S$ reconstruction. The efficiency
differences
$\Delta_{S\pm}=\Delta(\frac{\varepsilon_{S\pm}}{\varepsilon_{K\pi,S\pm}})$
of $\kpi$ between data and MC simulation are studied for the modes
$S\pm$. We use control samples to study $\Delta_{S\pm}$. The $\kpi$
final state is used for studying $\Delta_{S\pm}$ in the $K^+K^-$ and
$\pi^+\pi^-$ modes; $K^-\pi^+\pi^0$ is used for the $\pi^0\pi^0$,
$\rho\pi^0$, $K^0_S\pi^0$ and  $K^0_S\eta$  modes;
$K^-\pi^+\pi^0\pi^0$ is used for the $K^0_S\pi^0\pi^0$ mode; and
$K^+\pi^-\pi^-\pi^+$ is used for the $K^0_S\omega$  mode. We determine
$\Delta_{S\pm}$ in different $CP$-tag modes by comparing the ratio of
the DT yields to the ST yields between data and MC. We find that
$\Delta_{S\pm}$ are at 1\% level for different $CP$-tag modes. In the
formula of $\mathcal{A}^{CP}_{K\pi}$, the dependence of
$\Delta_{S\pm}$ on the $CP$ mode is not canceled out. The resulting
systematic uncertainty on $\mathcal{A}^{CP}_{K\pi}$ is
$0.2\times10^{-2}$.

Some systematics arise from effects which act among several $CP$
modes simultaneously. The efficiency of the cosmic and Bhabha veto
(only for the $KK$ and $\pi\pi$ modes) is studied based on the
inclusive MC sample. We compare the obtained $\mathcal{A}^{CP}_{K\pi}$
with and without this requirement and take the difference of $0.6\times10^{-3}$ as
a systematic uncertainty. For the $CP$ modes involving $K^0_S$,
$CP$-violating $K^0_L \to\pi^+\pi^-$ decays are also considered. Using
the known branching fraction, we find this causes the change on
$\mathcal{A}^{CP}_{K\pi}$ to be $0.8\times10^{-3}$.

Other systematic uncertainties, relevant to $\mathcal{A}^{CP}_{K\pi}$,
are listed in Table~\ref{uncor_sys}, which are uncorrelated among
different $CP$ modes.

The $\delE$ requirements are mode-dependent. We study possible biases
of our requirements by changing their values; we take the maximum
variations of the resultant $\br{D^{S\pm}\to K\pi}$ as systematic
uncertainties.

Fitting the $\mbc$ distributions involves knowledge of detector
smearing and the effects of initial-state and final-state
radiation. In the case of ST fits, we scan the smearing parameters
within the errors determined in our nominal fits. The maximum changes
to $n_{S\pm}$ are taken as a systematic uncertainty. For the DT fits,
we obtain checks on $n_{K\pi, S\pm}$ with one-dimensional fits to
$\mbc(S)$ with inclusion of floating smearing functions. The outcomes
of $\br{D^{S\pm}\to K\pi}$ are consistent with those determined from
the two-dimensional fits, and any small differences are treated as
systematic uncertainties.

Systematic effects due to the $CP$ purities are checked, as stated in
Sec.~\ref{CPdouble}. We introduce the $CP$ purities $f_S$ in
calculating the $\br{D^{S\pm}\to K\pi}$ under different $CP$ tagging
modes and obtain the corrected $\br{D^{S\pm}\to K\pi}$. We set the
lower limits of $f_S$ and take the corresponding maximum changes as
part of systematic uncertainties.

\section{Results}\label{result}

We combine the branching fractions $\br{D^{S+}\to K^-\pi^+}$ and $\br{D^{S-}\to K^-\pi^+}$  in Eq.~\eqref{22}  from two kinds of the $CP$ modes based on the
standard weighted least-square method~\cite{PDG}. Following
Eq.~\eqref{12}, we obtain
$\mathcal{A}^{CP}_{K\pi}=\ResultOfAcp$, where the first uncertainty is
statistical and the second is systematic. The mode-dependent
systematics are propagated to $\mathcal{A}^{CP}_{K\pi}$ and combined
with the mode-correlated systematics.
The values of $\mathcal{A}^{CP}_{K\pi}$ obtained for the 15 different $CP$ mode
combinations are also checked as listed in Table~\ref{A_Kpi}.  Within statistical uncertainties, they are consistent with each
other.

With external inputs of $r^2=(3.50\pm 0.04)\times10^{-3}$, $y=(6.7\pm0.9)\times10^{-3}$ from HFAG~\cite{HFAG} and $R_\mathrm{WS}=(3.80\pm0.05)\times10^{-3}$ from PDG~\cite{PDG}, $\cos\delta_{K\pi}$ is determined to be $\ResultOfCosStrPh$,
where the third uncertainty is due to the errors introduced from the external inputs.

\begin{table*}[tp!]
\caption{ Values of $\mathcal{A}^{CP}_{K\pi}$ in units of $10^{-2}$ extracted from the
  15 different combinations of CP decay modes.  The errors shown are
  statistical only.} {\label{A_Kpi}}
\begin{center}
\begin{tabular}{c*{5}{|c}}
\hline
\hline
\backslashbox{$CP-$}{$CP+$} &$K^+K^-$&$\pi^+\pi^-$&$K^0_S\pi^0\pi^0$ &$\pi^0\pi^0$&$\rho\pi^0$\\
\hline
$K^0_S\pi^0$ &$13.8\pm1.8$&$14.5\pm2.4$&$10.0\pm2.3$
&$8.0\pm3.7$&$12.2\pm2.0$\\
\hline
$K^0_S\eta$  &$15.5\pm3.5$&$16.3\pm3.9$&$11.8\pm3.8$
&$9.7\pm4.8$&$14.0\pm3.6$\\
\hline
$K^0_S\omega$ &$13.5\pm2.2$&$14.2\pm2.8$&$9.7\pm2.7$
&$7.7\pm3.9$&$11.9\pm2.4$\\
\hline
\hline
\end{tabular}
\end{center}
\end{table*}

\section{Summary}
We employ a $CP$ tagging technique to analyze a sample
  of 2.92\,fb$^{-1}$
quantum-correlated data of $\ee\to D^0\dbarzero$ at the $\psi(3770)$
peak. We measure the asymmetry $\mathcal{A}^{CP}_{K\pi} =
\ResultOfAcp$. Using the inputs of $r^2$ and $y$ from HFAG~\cite{HFAG}
and $R_\mathrm{WS}$ from PDG\cite{PDG}, we obtain $\cos\delta_{K\pi} =
\ResultOfCosStrPh$. The first uncertainty is statistical, the second
is systematic, and the third is due to the external inputs. Our result
is consistent with previous results from CLEO~\cite{CLEO-c2}. Our
result is the most precise to date, and  helps to constrain the
$\dzero$-$\dbarzero$ mixing parameters and the angle $\phi_3$
in the unitarity triangle of the CKM matrix.

\section{Acknowledgments}
The BESIII collaboration thanks the staff of BEPCII and the computing center for their strong support. This work is supported in part by the Ministry of Science and Technology of China under Contract No. 2009CB825200; National Natural Science Foundation of China~(NSFC) under Contracts Nos. 10625524, 10821063, 10825524, 10835001, 10935007, 11125525, 11235011, 11275266; Joint Funds of the National Natural Science Foundation of China under Contracts Nos. 11079008, 11179007; the Chinese Academy of Sciences~(CAS) Large-Scale Scientific Facility Program; CAS under Contracts Nos. KJCX2-YW-N29, KJCX2-YW-N45; 100~Talents Program of CAS; German Research Foundation DFG under Contract No. Collaborative Research Center CRC-1044; Istituto Nazionale di Fisica Nucleare, Italy; Ministry of Development of Turkey under Contract No. DPT2006K-120470; U. S. Department of Energy under Contracts Nos. DE-FG02-04ER41291, DE-FG02-05ER41374, DE-FG02-94ER40823, DESC0010118; U.S. National Science Foundation; University of Groningen (RuG) and the Helmholtzzentrum fuer Schwerionenforschung GmbH (GSI), Darmstadt; WCU Program of National Research Foundation of Korea under Contract No. R32-2008-000-10155-0.

\end{document}